\documentclass[12pt,preprint]{aastex}
\newcommand{\etal}{\mbox{\rm et al.~}}
\newcommand{\ms}{\mbox{m s$^{-1}~$}}
\newcommand{\ks}{\mbox{km s$^{-1}~$}}

\newcommand{\mse}{\mbox{m s$^{-1}$}}
\newcommand{\msun}{M$_{\odot}~$}

\newcommand{\msune}{M$_{\odot}~$}
\newcommand{\lsun}{L$_{\odot}~$}
\newcommand{\rsun}{R$_{\odot}~$}

\newcommand{\mjup}{M$_{\rm JUP}~$}

\newcommand{\msini}{$M \sin i~$}
\newcommand{\vsini}{$v \sin i~$}

\newcommand{\chisq}{$\sqrt{\chi_{\nu}^2}~$}

\newcommand{\teff}{$T_{\rm eff}~$}

\newcommand{\fe}{{\rm [Fe/H]}}
\newcommand{\logg}{${\rm \log g}~$}
\newcommand{\rhk}{$\log R^\prime_{HK}$}
\newcommand{\shk}{$S_{HK}$}
\newcommand{\prot}{$P_{ROT}~$}
  
\slugcomment{to appear in ApJ}
  
\shortauthors{Robinson \etal}
\shorttitle{New Planets}
\begin{document}
  
\title{Two Jovian-Mass Planets in Earthlike Orbits \altaffilmark{1}}
\author{Sarah E. Robinson\altaffilmark{2},
Gregory Laughlin\altaffilmark{2},
Steven S. Vogt\altaffilmark{2},
Debra A. Fischer\altaffilmark{3},
R. Paul Butler\altaffilmark{4},
Geoffrey W. Marcy\altaffilmark{5},
Gregory W. Henry\altaffilmark{6},
Peter Driscoll\altaffilmark{3,7},
Genya Takeda\altaffilmark{8},
John A. Johnson\altaffilmark{5}}
  
\email{ser@ucolick.org}
  
\altaffiltext{1}{Based on observations obtained at the W. M. Keck
Observatory, which is operated by the University of California and the
California Institute of Technology. Keck time has been granted by NOAO
and NASA.}

\altaffiltext{2}{UCO/Lick Observatory, University of California at Santa
Cruz, Santa Cruz, CA 95064}

\altaffiltext{3}{Department of Physics \& Astronomy, San Francisco State
University, San Francisco, CA  94132; fischer@stars.sfsu.edu}

\altaffiltext{4}{Department of Terrestrial Magnetism, Carnegie Institute
of Washington DC, 5241 Broad Branch Rd. NW, Washington DC, USA
20015-1305}

\altaffiltext{5}{Department of Astronomy, University of California,
Berkeley, CA USA 94720}
  
\altaffiltext{6}{Center of Excellence in Information Systems, Tennessee
State University, 3500 John A. Merritt Blvd., Box 9501, Nashville, TN
37209}

\altaffiltext{7}{Department of Earth and Planetary Sciences, Johns
Hopkins University, Baltimore, MD 21218}

\altaffiltext{8}{Department of Physics and Astronomy, Northwestern 
University, 2145 Sheridan Road, Evanston, IL 60208}

\begin{abstract}
We report the discovery of two new planets: a 1.94 \mjup planet in a
1.8-year orbit of HD 5319, and a 2.51 \mjup planet in a 1.1-year orbit
of HD 75898.  The measured eccentricities are 0.12 for HD 5319~b and
0.10 for HD 75898~b, and Markov Chain Monte Carlo simulations based on
the derived orbital parameters indicate that the radial velocities of
both stars are consistent with circular planet orbits.  With low
eccentricity and $1 < a < 2$~AU, our new planets have orbits similar to
terrestrial planets in the solar system.  The radial velocity residuals
of both stars have significant trends, likely arising from substellar or
low-mass stellar companions.
\end{abstract}


\section{Introduction}

Doppler searches for Jupiter-mass planets are nearly complete for
semimajor axes $0.03 \leq a \leq 3$~AU.  The distribution of exoplanet
semimajor axes within this range reveals a paucity of planets orbiting
less than $0.5$~AU from their host stars (Butler \etal 2006).  Instead,
many giant planets are being found on terrestrial planet-like orbits, in
or near the habitable zone: 25\% of 212 known exoplanets within 100 pc
have semimajor axes between 1.0 and 2.0 AU\footnote{See catalog at
www.exoplanets.org.}.  These observations suggest that planetary systems
often have habitable zones dominated by gas giants.  In this paper, we
announce the discovery of Jupiter-mass planets orbiting HD 5319 and HD
75898.  Both planets have nearly circular orbits with semimajor axes
between 1 and 2 AU.

HD 5319 and HD 75898 were selected for the Keck planet search after
being flagged as metal-rich by the N2K consortium, on the basis of
photometry and low-resolution spectroscopy (Ammons \etal 2006, Robinson
\etal 2007).  The N2K project's primary goal was to identify metal-rich
stars likely to host hot Jupiters, which have high transit probabilities
(Fischer \etal 2005).  So far, one transiting hot Saturn (Sato \etal
2005) and six planets with periods $P < 15$ days (Wright \etal 2007,
Johnson \etal 2006, Fischer \etal 2006, and Fischer \etal 2005) have
been discovered among the N2K targets.  However, the planet-metallicity
correlation holds for all orbital periods, making the N2K target list a
good source for discoveries of longer-period planets as well.  The new
discoveries reported in this paper, HD 5319~b and HD 75898~b, are two of
the seven intermediate-period planets so far found orbiting N2K target
stars (see also Wright \etal 2007, Fischer \etal 2007).

In \S 2, we report our observations and Keplerian fit to HD 5319.  We
discuss the HD 75898 system in \S 3.  In \S 4, we discuss the implied
presence of long-period stellar or substellar companions orbiting each
star.  We present discussion and conculsions in \S 5.

\section{HD 5319}

\subsection{Stellar Characteristics}

HD 5319 is a subgiant with $M_V=3.05$, $V=8.05$, $B-V=0.985$, and
Hipparcos parallax (ESA 1997) of $0.010 ''$, corresponding to a distance
of 100 parsecs.  High-resolution spectroscopic analysis (Valenti \&
Fischer 2005) yields \teff = 5052 $\pm$ 50K, \logg = 3.57 $\pm$ 0.15,
\vsini = 3.31 $\pm$ 0.50 \ks, and \fe = 0.15 $\pm$ 0.05 dex.  HD 5319's
spectral type is listed as K0 III in the SIMBAD database and as G5 IV in
the Hipparcos catalog.  The star's $M_V$ and \logg values are most
consistent with the G5 IV designation.

The luminosity is 4.6 $L_{\odot}$, including a bolometric correction of
$-0.259$ (VandenBerg \& Clem 2003). The luminosity and effective
temperature imply a stellar radius of 2.8 $R_{\sun}$. We estimate
stellar masses using theoretical stellar models based on the Yale
Stellar Evolution Code as described in Takeda \etal (2007).  The fine
grid of evolutionary tracks have been tuned to the uniform spectroscopic
analysis of Valenti \& Fischer (2005) and provide posterior
distributions for stellar mass, radius, gravity and age.  Based on this
analysis, we derive a stellar mass of $1.56 \: M_{\odot}$, a radius of
$3.26 \: R_{\odot}$, higher than implied by the bolometric luminosity,
and an age of 2.4~Gyr for this subgiant.  As a measure of uncertainty,
the lower and upper 95\% confidence intervals are provided in
parentheses in Table 1 for the stellar mass, age and radius.

Measurement of the core of the Ca H\&K lines (Figure 1) show
that the star is chromospherically inactive.  From 30 observations, we
measure mean values of the Ca H\&K indices of \shk = 0.12 and \rhk =
-5.34.  Based on the values of \shk~and \rhk, we derive a rotational
period of \prot = 19.0 days (Noyes \etal 1984).  However, we caution
that the interpretation of \shk~and \rhk and their correlation with
\prot may be subject to systematic errors for evolved stars, since the
\prot calibration was created for main-sequence stars.

We also monitored the star's brightness with the T10 0.8~m automatic
photometric telescope (APT) at Fairborn Observatory (Henry 1999, Eaton
\etal 2003). The T10 APT measures the brightness of program stars
relative to nearby constant comparison stars with a typical precision of
0.0015--0.0020 mag for a single measurement.  We obtained 89 Str\"omgren
$b$ and $y$ brightness measurements spanning 438 days between 2004
October and 2006 January.  The standard deviation of a single
observation from the mean was 0.0017 mag, comparable to the measurement
precision, which provides an upper limit to photometric variability in
HD~5319.  A periodogram analysis found no significant periodicity
between 1 and 220 days.  Thus, our photometry confirms the star's
low level of chromospheric activity.

\subsection{Doppler Observations and Keplerian Fit}

Doppler observations were made at the Keck telescope using HIRES (Vogt
et al. 1994) with an iodine cell to model the instrumental profile and
to provide the wavelength scale (Butler \etal 1996). An exposure meter
maintains a constant signal-to-noise ratio of about 200 in our spectra,
yielding relatively uniform radial velocity precision.  We obtained 30
observations of HD 5319.  The observation dates, radial velocities and
measurement uncertainties are listed in Table 2 and plotted in Figure 2.

The periodogram of the radial velocities (Figure 3) shows a strong,
broad peak in the power spectrum, spanning 600-900 days.  This peak is
wide because of modest phase sampling for HD 5319~b.  To estimate the
False Alarm Probability (FAP), the probability that the power in the
highest peak is an artifact of noisy data or timing of observations, we
use the bootstrap Monte Carlo method of Cumming (2004).  We generated
10,000 data sets of noise using the measured stellar velocities,
selected with replacement from residuals about the mean velocity, and
calculated the periodogram for each synthetic RV data set. The fraction
of trials with maximum periodogram power that exceeds the observed value
gives the FAP (Cumming 2004).  Figure 4 shows a histogram of the tallest
peak height in each trial periodogram.  Only 13 of 10,000 synthetic data
sets yielded any peak with higher power than in the true periodogram,
for ${\rm FAP} = 0.0013$ (Table 3).  The probability that the 600-900
day periodogram peak arises from a true physical source is therefore
$99.87\%$, suggesting this period range should be searched for a
Keplerian orbital fit.

The final task before determining the orbit of HD 5319~b is to assess
the astrophysical sources of error in radial velocity measurements.  In
addition to velocity errors arising from our measurement uncertainties
(including photon shot noise), the star itself can have cool spots,
granular convective flows, or $p$-mode oscillations that contribute
non-dynamical velocity noise.  These noise sources are collectively
termed ``jitter''.  For purposes of fitting a Keplerian model, the
stellar jitter is added in quadrature to the formal instrumental errors.
Jitter is not included in the tabulated measurement uncertainties for
the radial velocity sets.

We empirically estimate stellar jitter based on the chromospheric
activity of the star and spectral type, following Wright (2005).  The
20$^{th}$ percentile, median, and 80$^{th}$ percentile jitter amplitudes
of stars at the chromospheric activity level and evolutionary stage of
HD 5319 are 4.6~\mse, 5.7~\mse, and 9.5~\mse, respectively.  We adopt
the 20$^{th}$ percentile value as a conservative jitter estimate (Table
1).  The $p$-mode oscillation component of the jitter is $\sim
0.9$~\mse, according to the solar scaling relation of Kjeldsen \&
Bedding (1995).

A Levenberg-Marquardt (LM) fitting algorithm was used to model the
radial velocities of HD 5319. The best-fit Keplerian model gives an
orbital period of 674.6 $\pm$ 16.9 d, with semi-velocity amplitude $K =
33.6 \pm$ 4.3 \mse, and orbital eccentricity $e = 0.12 \pm 0.08$.  We
include a center of mass acceleration $dv/dt = 9.11$~\ms~yr$^{-1}$,
corresponding to a linear trend in the residual radial velocities.  The
best fit has RMS~=~6.08~\ms and \chisq~=~1.22, including 4.6~\ms for
astrophysical jitter.  Adopting a stellar mass of 1.56~$M_{\odot}$, we
derive \msini~=~1.94 \mjup and semimajor axis $a = 1.75$~AU (angular
separation, $\alpha = 0.'' 0175$).  The orbital solution is listed in
Table 3 and the RV data are plotted with the best-fit Keplerian model
(solid line) in Figure 2.

Uncertainties in the orbital parameters are first estimated with a
model-based bootstrap Monte Carlo analysis.  First, we find the best fit
Keplerian model.  Then, for each of 250 trials, that theoretical best
fit is subtracted from the observed radial velocities.  The residual
velocities are then scrambled (with replacement) and added back to the
theoretical best fit velocities and a new trial Keplerian fit is
obtained. We adopt the standard deviation of each orbital parameter for
the 250 Monte Carlo trials as the parameter uncertainty.  The
uncertainties of the Keplerian parameters of HD 5319~b are listed in
Table 3.

In order to confirm the orbital parameters of HD 5319~b, a Markov Chain
Monte Carlo (MCMC) simulation was carried out for the HD 5319
velocities.  This analysis, which gives posterior probability
distribution for the orbital parameters, can be a useful check of the
convergence of the Levenberg-Marquardt fitting algorithm, particularly
when the modeled \chisq space is confused with several local minima.
For example, poor phase coverage might result in an aliased value of the
period.  A bimodal MCMC posterior distribution for the period would
indicate the need for more observations to break the degeneracy.

Posterior probability distributions of $P$, $e$ and $K$ are shown for HD
5319~b in Figure 5.  Because the orbit is nearly circular, the time of
periastron passage and longitude of periastron are not well constrained,
and the MCMC histograms are nearly flat.  This ambiguity is also
reflected in the large uncertainties ($\sim 1/8$~orbit) for ${\rm
T}_{\rm p}$ and $\omega$ inferred from the orbit-based bootstrap
simulations.  The eccentricity distribution has mean $e = 0.09$, with
our reported value of $e = 0.12$ lying within $1 \sigma$ of the mean.
The mean of the period distribution, 686~days, is consistent with the
period determined by the LM analysis, 675~days.  The MCMC simulations
settle on a somewhat larger value of velocity semiamplitude, $K =
39$~\ms, than the LM analysis ($K = 33.8$~\mse), a difference of
$1.4 \, \sigma$.

To assess the wisdom of adding the extra term $dv/dt$ to our fit, we
perform an $F$-test for an additional term (Bevington \& Robinson 1992).
We define $\Delta \chi^2$ as the difference in unreduced $\chi^2$
between the best fits obtained with and without the $dv/dt$ term, and
$\chi^2_{\nu}$ as the reduced $\chi^2$ of the published fit, including
$dv/dt$.  The quantity
\begin{equation}
F = {\Delta \chi^2 \over \chi^2_{\nu}} 
\label{fstat}
\end{equation}
follows an $F$-distribution with one numerator degree of freedom and
$\nu = N_{\rm obs} - 7$ denominator degrees of freedom, where $N_{\rm
obs}$ is the number of observations (30 for HD 5319).  The best fit
obtained without the $dv/dt$ term has $\chi^2 = 61.4$ (\chisq = 1.60),
giving $F = 18.2$.  The probability $P(F;1,23)$ that a randomly selected
F exceeds this value is 0.00029, for a less than 1 in 1,000 chance that
the fit improvement provided by the $dv/dt$ term is spurious.
Therefore, there is strong evidence that a long-period companion is
accelerating the center of mass of the HD 5319~a-b system.

Noticing a smooth variation in the residuals of a one-planet fit,
one might be tempted to fit a second Keplerian with a longer period.
However, in the case of HD 5319, this is premature: the linear
correlation coefficient of the one-planet residuals is 0.85, indicating
that a linear model describes the variation in these residuals well.  We
do not yet detect any curvature in the radial velocity signature of the
second companion, so we refrain from fitting a full Keplerian or a
circular orbit.  We allow the period of this long-period companion to
remain undetermined, and approximate its effects on the system with a
constant acceleration $dv/dt$.

\section{HD 75898}

\subsection{Stellar Characteristics}

HD 75898 has $V = 8.03$, $B-V = 0.626$, and Hipparcos parallax (ESA
1997) of $0.012 ''$, corresponding to a distance of 80.6 parsecs.
Spectroscopic analysis yields \teff = 6021 $\pm$ 50K, \logg = 4.16 $\pm$
0.15, \vsini = 4.54 $\pm$ 0.50 \ks, and \fe = 0.27 $\pm$ 0.05 dex.  The
absolute visual magnitude is $M_V = 3.49$, and the luminosity is
3.0~\lsun (with bolometric correction of -0.039).  Although the SIMBAD
spectral type designation is G0V, the luminosity, temperature and
surface gravity are more consistent with a metal-rich F8V star.  The
value of $M_V$ indicates that the star is just beginning to evolve onto
the subgiant branch.  From the stellar luminosity and surface gravity,
we derive a stellar radius of $1.6 R_{\sun}$, identical to the radius
derived from evolutionary tracks.  A stellar mass of 1.28~\msun and an
age of 3.8~Gyr are derived from evolutionary tracks (Takeda \etal 2007).
The physical parameters of HD 75898 are listed in Table 1.

HD 75898 was selected for the Keck planet search after being observed by
the N2K low-resolution spectroscopic survey (Robinson et al. 2007),
carried out at the 2.1m telescope at KPNO from August 2004 to April
2005.  The atmospheric parameters measured from N2K spectra were $T_{\rm
eff} = 5983 \pm 82$~K, ${\rm [Fe/H]} = 0.22 \pm 0.07$~dex, and $\log \;
g = 4.22 \pm 0.13$~dex.  These values agree with the Keck measurements
within uncertainties.

Figure 1 shows that the star is chromospherically inactive, with no
observed emission in Ca II H\&K . We derive mean \shk = 0.15 and \rhk =
-5.02, with a corresponding rotational period \prot = 12.6 d.  The
caution that the \prot measurement may be affected by systematic errors
for evolved stars applies to HD 75898 as well, since this star is
beginning to move off the main sequence.  Wright (2005) reports
20$^{th}$ percentile, median, and 80$^{th}$ percentile jitter amplitudes
of 2.6~\mse, 4.0~\mse, and 6.2~\mse, for stars with similar activity
level and evolutionary stage to HD 75898.  Again, we adopt a
conservative, 20$^{th}$ percentile jitter estimate of 2.6~\ms (Table 1).
The $p$-mode oscillation component of the jitter is $\sim 0.5$ \ms
(Kjeldsen \& Bedding 1995).  The stellar characteristics are summarized
in Table 1.

\subsection{Doppler Observations and Keplerian Fit}

We obtained 20 observations of HD 75898.  Observation dates, radial
velocities and instrumental uncertainties in the radial velocities (not
including stellar jitter) are listed in Table 4.  The periodogram for
this data set (Figure 7) shows a strong peak at 446 days.  Once again,
we calculate the FAP by sampling the observed radial velocities with
replacement, keeping the original observation times, and calculating the
maximum periodogam power for the scrambled velocities.  In 10,000
synthetic data sets, no periodogram had higher power than the original
446-day peak.  The FAP for this peak is $< 0.0001$ (Table 3), indicating
a better than 0.9999 probability that the periodicity in radial
velocities has an astrophysical source, and is not caused by noise.  The
histogram of periodogram power in the tallest periodogram peak in each
of 10,000 trials is plotted in Figure 8.

There is also a peak in the periodogram at 200 days, which may be an
alias of the true, $\sim 400$-day period; an artifact of the $1/2$-year
observing season of HD 75898, which is near the ecliptic.  Two other
possible explanations for the 200-day peak are that it arises from the
modest eccentricity ($e \approx 0.1$) of the best-fit 418-day orbit, or
that there is a second planet in the system with a period near 200 days.
The observations between 2004 January and 2006 May, which do not include
the minimum of the radial velocity curve, can be modeled credibly with
Keplerian orbits of either $\sim 200$ or $\sim 400$ days.  However, when
the four most recent observations, which do cover the radial-velocity
minimum, are included, the degeneracy is broken and single planets with
200-day orbits do not fit the data.

The best-fit Keplerian model gives a period of 418.2 $\pm$ 5.7 days,
with semi-velocity amplitude $K = 58.2 \pm 3.1$ \mse, and orbital
eccentricity $e = 0.10 \pm 0.05$.  The RMS to the fit is 5.48 \ms with
\chisq = 1.77, including the estimated astrophysical jitter of 2.6 \mse.
Adopting a stellar mass of 1.28 \msune, we derive \msini = 2.51 $M_{\rm
JUP}$.  The corresponding semimajor axis is $a = 1.19$~AU, and the
angular separation is $\alpha = 0.'' 0148$.  The residual velocities
show a strong trend, $dv/dt = -14.6$ \ms yr$^{-1}$, suggesting that an
additional companion orbits the star.  The Keplerian orbital parameters
are listed in Table 3 and plotted with the best-fit Keplerian model
(solid line) in Figure 6.

To assess whether the constant acceleration $dv/dt$ should be included
in the fit, we again perform the $F$-test for an additional term given
in Equation \ref{fstat}.  The best-fit Keplerian without the $dv/dt$
term has $\chi^2 = 142.5$ (\chisq = 3.19).  The $F$-statistic comparing
the best fits with and without $dv/dt$ is 32.5.  There are 20
observations of this star, giving the $F$ distribution 13 denominator
degrees of freedom.  The probability $P(F;1,13)$ that the fit
improvement from including $dv/dt$ is spurious is only $7.3 \times
10^{-5}$. The detected acceleration of the HD 75898~a-b center of mass
is therefore almost certainly real, and not an artifact of noise.
Further evidence for a long-period companion to HD 75898 is provided by
the periodogram, which rises toward a 2000-day period (almost twice the
length of our observational baseline).  The correlation coefficient for
a linear fit to the single-planet residuals is $r = -0.96$, indicating
that variation in RV residuals is well described by a constant
acceleration.  The relatively sparse sampling precludes detection of
curvature from any additional planets at this time.

We carried out a Markov Chain Monte Carlo simulation for the radial
velocity residuals of HD 75898.  The resulting posterior distributions
for period, radial velocity semi-amplitude, and eccentricity are shown
in Figure 9.  For this low-eccentricity orbit, time of
periastron passage and longitude of periastron are not well constrained.
The fact that the MCMC eccentricity distribution peaks at zero suggests
that the orbit of HD 75898~b could in fact be circular.  The mean
eccentricity in the MCMC posterior distribution is $0.1$, in agreement
with the Levenberg-Marquardt value of $0.10 \pm 0.05$.  The mean of the
MCMC period distribution is 417~days, which is well matched with the
results of the LM analysis ($P = 418$~days).  For velocity
semi-amplitude $K$, the MCMC results also reproduce the LM results, with
mean $K = 58$~\ms.

Being near the ecliptic ($\delta = 33\degr$), with a period near one
year (418 days), HD 75898 presents a special hazard for planet
detection.  The observing season for HD 75898 is just 7 months, so only
half of the orbital phase is visible during one year.  At the same time,
the visible phase of HD 75898~b's orbit advances only $12\%$ per year.
Although our observational baseline covers 3 years and 2/3 of the orbit,
it would take 5 years to obtain full phase coverage.  We expect the
orbital solution to be revised as more observations of HD 75898 are
obtained.

Observations near periastron passage contain the most information about
the orbit, particularly eccentricity (e.g. Endl \etal 2006).  If our
best-fit orbit is correct, we have observed the periastron passage to
within 4 days (JD 2453747).  This fact, combined with the results of the
MCMC simulation for eccentricity, leads us to believe that our basic
discovery is correct, that HD 75898 has a planet with a minimum mass
of 2~\mjup in a nearly circular orbit near 1 AU.

\section{Long-Period Companions}

The radial velocity residuals of HD 5319 and HD 75898 have significant
linear trends, $|dv/dt| \geq 9$~\ms yr$^{-1}$.  This indicates that the
center of mass of each two-body system is accelerating, which cannot
happen unless there is a third component in each system.  Both stars,
then, show evidence of long-period companions with incomplete phase
coverage during our observational baseline.  The possibility of finding
brown dwarfs orbiting sunlike stars is a tantalizing one, warranting
further analysis of the one-planet residuals of HD 5319 and HD 75898.
Brown dwarf companions might reside in the brown dwarf desert (McCarthy
\& Zuckerman 2004, Grether \& Lineweaver 2006), the dearth of substellar
companions to main-sequence stars with $a < 1200$~AU.  Another
possibility is that the third components are giant planets with $P \ga
2000$ days.  Even the presence of stellar companions would make HD 5319
and HD 75898 unusual planet hosts, as new evidence indicates planet
occurrence is infrequent in binaries closer than 120~AU (Eggenberger \&
Udry 2007).  In this section, we analyze possible configurations of the
HD 5319 and HD 75898 systems.

The possible companion types---planets, brown dwarfs, and stars---are
restricted to a particular semimajor axis range by the measured $dv/dt$
and the long-term dynamical stability of each system.  Although these
ranges overlap substantially when all potential variations in time of
periastron passage, line of apsides and eccentricity are taken into
account, the general pattern is $a_* \ga a_{\rm bd} \ga a_{\rm planet}$.
This pattern can be illustrated by the simple example of a circular
orbit: the star reaches radial velocity semiamplitude when the planet
has moved 1/4 orbit from its ephemeris, so we can calculate the
approximate semiamplitude by
\begin{equation}
K \approx \left ( {P \over 4} \right ) {dv \over dt}.
\label{perk}
\end{equation}
Equation \ref{perk} shows that the longer a star maintains the measured
constant $dv/dt$, the higher the mass of the companion.  For eccentric
orbits, the proportionality constant relating $K$ and $P$ changes, but
the pattern $a_* \ga a_{\rm bd} \ga a_{\rm planet}$ holds.

The smallest possible semimajor axis for component c, $a_{\rm min}$, is
determined by the requirement that the two companions in each system do
not experience close encounters, which could lead to large perturbations
of both orbits.  Absent any dynamical considerations, $a_{\rm min}$
would correspond to a highly eccentric orbit with the apoastron passage
near the midpoint of our observations, and with a period only slightly
longer than our time baseline.  However, the more eccentric the outer
component's orbit, the nearer its approach to the inner planet during
periastron passage.

Assuming nonresonant systems, we can set a lower limit to the distance
of closest approach between components b and c.  David \etal (2003)
examine of the stability of a two-planet system, an intermediate-mass
companion exterior to an Earth-mass planet on a circular orbit at 1~AU.
They find that an outer planet with mass $1 M_{\rm JUP}$ and $R_{\rm
peri} \sim 2.5$~AU give a mean ejection time of 1~Gyr for the
terrestrial planet.  Although HD 5319 and HD 75898 have far more massive
inner planets than the theoretical system of David et al., we apply
their analysis because of the similar orbits of the inner planets in all
three systems.  Adopting the 1~Gyr stability criterion, we set the
minimum periastron distance of the c component as $R_{\rm min} \geq 2.5
\, a_{\rm inner}$.  For each component type (planet, brown dwarf, star),
the orbit corresponding to $a_{\rm min}$ must obey this stability
criterion and reproduce the observed $dv/dt$ within uncertainties.

In this section, we refer to HD 5319~c, HD 75898~c and ``the c
components.''  We are using this nomenclature as shorthand for ``implied
long-period companion,'' and are not claiming actual detections of
these objects.

\subsection{Planet Orbits}
\label{porbs}

If HD 5319~c or HD 75898~c is a planet, $a_{\rm min}$ is simply the
stability limit $R = 2.5 \, a_{\rm inner}$, and the orbit associated
with $a_{\rm min}$ is circular.  We note that giant planets on circular
orbits can have semimajor axis ratios less than 2.5, as Jupiter and
Saturn do, but $a_{\rm outer} / a_{\rm inner} < 2$ is rare among
exoplanets (Butler et al. 2006).  To find the minimum planet
mass for HD 5319~c and HD 75898~c, we substitute test values of $M \,
\sin i$ into the equations
\begin{equation}
\left ( {a \over {\rm AU}} \right )^3 = \left ({M_{\star} + M \, \sin i
\over M_{\odot}} \right ) \left ({P \over {\rm yr}} \right )^2,
\label{kepler}
\end{equation}
\begin{equation}
M \, \sin i = K \sqrt{1-e^2} \left [ {P (M_{\star} + M \, \sin i)^2
\over 2 \pi G} \right ]^{1/3},
\label{msini}
\end{equation}
and use the resulting value of $K$ to calculate a radial velocity curve.
$M_{\rm min,planet}$ is the lowest value of $M \, \sin i$ for which the
radial velocity slope, determined from a linear fit, matches the
observed $dv/dt$ within the uncertainties reported in Table 3:
$|dv/dt_{\rm calc} - dv/dt_{\rm obs}| \leq \sigma(dv/dt)$.

The minimum mass of HD 5319~c is $1.0 \; M_{\rm JUP}$.  This planet
would reside at $a = 4.4$ AU, and have a period of 2675 days (7.3 yr).
Figure 10 shows the radial velocity curve corresponding to this orbit,
together with the observed trend in the fit residuals of HD 5319~b.  HD
75898~c also has a minimum mass $M_{\rm min} = 1.0 \: M_{\rm JUP}$, in a
circular orbit with semimajor axis $a = 3.0$ AU and $P = 1656$ days (4.5
yr).  The resulting radial velocity curve, plus the measured trend in
the HD 75898~b residuals, are shown in Figure 11.  These orbital
solutions show that HD 5319~c and HD 75898~c could be similar, in
mass, semimajor axis and perhaps equilibrium temperature, to Jupiter.

For the maximum possible planet mass of HD 5319~c or HD 75898~c, we
adopt the IAU criterion that a planet does not burn deuterium, and so
$M_{\rm max} = 13 \: M_{\rm JUP}$ (Boss et al. 2007).  We can calculate
$a_{\rm max}$ and $P_{\rm max}$ for this borderline planet by examining
the limiting case where the periastron passage, which coincides with
ephemeris, occurs at the midpoint of our observations.  This is the part
of the radial velocity curve that varies most rapidly.  In principle,
$a_{\rm max}$ and $P_{\rm max}$ could become arbitrarily large as $e
\rightarrow 1$.  We adopt the convention of Patel et al. (2007) and
define $e_{\rm max} = 0.8$, since $90\%$ of spectrosopic binaries with
$P > 10$ yr have $e < 0.8$ (Pourbaix et al. 2004).

To calculate $a_{\rm max}$ for planet orbits, we substitute test values
of $a$ into equations \ref{kepler} and \ref{msini}, and use the
resulting value of $K$ to calculate a radial velocity curve.  We then
find the maximum $a$ for which $|dv/dt_{\rm calc} - dv/dt_{\rm obs}|
\leq \sigma(dv/dt)$.  For HD 5319~c, $a_{\rm max} = 85$ AU,
corresponding to a period of 625 yr.  For HD 75898~c, $a_{\rm max} = 65$
AU and $P_{\rm max} = 460$ yr.  In practice, it is extremely unlikely
that either object is a planet on this type of orbit: the probability of
catching these long-period, eccentric orbits exactly at periastron
passage is quite low.  The ranges of possible planet orbits for HD
5319~c and HD 75898~c are summarized in Table 5.

\subsection{Brown Dwarf Orbits}
\label{bdorbs}

To find $a_{\rm min}$ for brown dwarf orbits, we examine the limiting
case where apoastron coincides with ephemeris at the midpoint of our
observational baseline.  This configuration gives the RV curve that most
nearly approximates a straight line.  Recalling that high mass implies
high semimajor axis (c.v. Equation \ref{perk}), we set
\msini~=~13~\mjup, the minimum possible brown dwarf mass.  Substituting
test values of $a$ and $e$ into Equations \ref{kepler} and \ref{msini},
we find the semiamplitude $K$ for each $a$, $e$ pair that meets the
stability criterion $a \, (1-e) \geq 2.5 \, a_{\rm inner}$.  The linear
slopes of the resulting radial velocity curves are examined to find the
minimum $a$ where $\Delta(dv/dt) \leq \sigma(dv/dt)$ (Table 3).  For HD
5319~c, $a_{\rm min} = 9.8$ AU for a brown dwarf.  This orbit has $e =
0.55$ and $P = 24$ yr.  If HD 75898~c is a brown dwarf, $a_{\rm min} =
7.5$ AU and $P_{\rm min} = 18$ yr, with eccentricity $e = 0.60$.

We determine $a_{\rm max}$ for brown dwarf orbits by setting
\msini~=~83.8 \mjup, the minimum mass for hydrogen fusion.  We follow
the same method outlined in \S \ref{porbs} for finding $a_{\rm max}$,
once again assuming $e_{\rm max} = 0.8$.  For HD 5319~c, $a_{\rm max} =
190$ AU and $P_{\rm max} = 2045$ yr for brown dwarf orbits.  For HD
75898~c, $a_{\rm max} = 170$ AU and $P_{\rm max} = 1900$ yr.

We note that the brown dwarf semimajor axis ranges implied by our
measured $dv/dt$, $9.8 \la a \la 190$ AU for HD 5319~c and $7.5 \la a
\la 170$ AU for HD 5319~c, fall directly in the brown dwarf desert
(McCarthy \& Zuckerman 2004).  The ranges of possible brown dwarf
periods and semimajor axes for HD 5319~c and HD 75898~c are summarized
in Table 5.

\subsection{Star Orbits}
\label{storbs}

To find $a_{\rm min}$ for stellar companions to HD 5319 and HD 75898, we
set \msini~=~83.8~\mjup, the minimum mass for a main-sequence star.
Once again, we set the apoastron passage to coincide with the ephemeris
and place it at the midpoint of our observations, finding the best
approximation to a linear RV curve.  We follow the procedure outlined in
\S \ref{bdorbs}, testing a grid of $a$ and $e$ values which meet the
stability criterion and finding the minimum $a$ for which the linear RV
slope matches our measured $dv/dt$ within uncertainties (Table 3).  The
minimum semimajor axis for HD 5319~c, if it is a star, is $a_{\rm min} =
22$ AU, with $P_{\rm min} = 81$ yr and $e = 0.8$.  For HD 75898~c,
$a_{\rm min,\star} = 17$ AU, $P_{\rm min,\star} = 58$ yr and $e = 0.8$.

We determine the maximum masses of stellar companions to HD 5319 and HD
75898 by noting that neither star was identified as a double-lined
spectroscopic binary (SB2) in the Keck spectra.  The minimum flux ratio
for detecting SB2s with HIRES is $\sim 0.01$.  This limit gives $M_V >
8.05$ HD 5319~c and $M_V > 8.49$ HD 75898~c.  The corresponding masses
are \msini~=~0.65 \msun and \msini~=~0.6 \msun, respectively (Yi et al.
2001).  Assuming periastron passage and ephemeris fall at the midpoint
of our observations and $e_{\rm max} = 0.8$, HD 5319~c has $a_{\rm max}
= 630$ AU and $P_{\rm max} = 10600$ yr.  HD 75898~c has $a_{\rm max} =
470$ AU and $P_{\rm max} = 7400$ yr.  The possible stellar orbits for HD
5319~c and HD 75898~c are summarized in Table 5.  Note that these orbit
determinations are extremely uncertain, as we are using a 3-yr
observational baseline to characterize orbits in the $10^2 - 10^4$-year
range.

\section{Discussion and Conclusions}

We have discovered two Jovian-mass planets in Earthlike orbits, $1 < a <
2$~AU, orbiting the stars HD 5319 and HD 75898.  Target selection of
both stars was performed by the N2K Consortium (Fischer et al. 2005).
For HD 75898, which was observed as part of the N2K low-resolution
spectroscopic survey (Robinson et al. 2007), we find good agreement
between the N2K and Keck atmospheric parameter estimates.

At 1.56~\msun, HD 5319 is on the verge of being a ``retired'' A-star
($M_{\star} > 1.6$~\msun) of the type discussed by Johnson et al.
(2007).  In all 9 previously known former A-dwarf planetary systems, the
planets orbit at semimajor axes $a \geq 0.78$~AU.  HD 5319~b fits this
pattern well, with $a = 1.75$~AU.  Although the total number of known
planet hosts with $M > 1.6$~\msun is small, Johnson et al. concluded
that the dearth of short-period planets around these stars is real, and
the semimajor axis distributions of planets orbiting intermediate-mass
and low-mass stars are different.  Furthermore, engulfment by the
expanding subgiant can only explain the disappearance of planets
orbiting at $a < 30 R_{\odot}$.  Burkert \& Ida (2007) point out that
the lack of short-period planets orbiting intermediate-mass stars can be
explained if these stars' protostellar disks have a shorter depletion
timescale than their low-mass counterparts.

Among orbits larger than the tidal circularization cutoff of 0.1 AU,
circular orbits, while not rare, are certainly not preferred.  Butler et
al. (2006) report that the distribution of eccentricities is nearly
uniform beyond 0.3 AU.  However, Meschiari et al. (2007, submitted)
performed a blind experiment where they presented users of the
Systemic\footnote{http://oklo.org} radial velocity-fitting console with
synthetic radial velocity data sets drawn from circular orbits.  The
recovered eccentricity distributions had median values between 0.1 and
0.2, indicating a bias toward finding eccentric orbits.  If the median
exoplanet eccentricity has been skewed higher than its true value by the
planet discovery process, solar system-like orbits, which seem
noteworthy in the context of so many eccentric exoplanets, may be quite
common.  With $1 < a < 2$ AU and $e \approx 0.1$, HD 5319~b and HD
75898~b have orbits quite similar to our own terrestrial planets.

HD 5319 and HD 75898 have radial velocity residuals that imply
additional companions in the system.  To account for our measured
center-of-mass accelerations, HD 5319~c and HD 75898~c must both be at
least 1~\mjup.  If the periods and masses of these objects are near the
minimum values recorded in Table 5, further radial velocity observations
might add these stars to the known list of multiple-planet systems
within a few years.  However, it is likely that these objects have
periods too long for radial-velocity follow-up.  In that case, HD 5319
and HD 75898 are good candidates for high-resolution imaging.  The
NIRC-2 coronagraph spotsize is $0.'' 5$, which would restrict the
detection space to $a > 50$~AU for HD 5319 and $a > 40$~AU for HD 75898.
With the NIRC-2+AO limiting contrast ratio of $0.1\%$, this detection
space includes massive brown dwarfs and low-mass stars.  The analytical
work of Matzner \& Levin (2005) supports the hypothesis that
protostellar disk fragmentation is not a viable formation mechanism for
star-brown dwarf binary pairs.  HD 5319 and HD 75898 could therefore
serve as laboratories for investigating the presumably rare phenomenon
of brown dwarf formation in protostellar disks.

\acknowledgements

SER thanks Eugenio Rivera and Peter Bodenheimer for helpful input on
this work.  We gratefully acknowledge the dedication and support of the
Keck Observatory staff, in particular Grant Hill for support with HIRES.
We thank the NASA and UC Telescope assignment committees for generous
allocations of telescope time.  The authors extend thanks to those of
Hawaiian ancestry on whose sacred mountain of Mauna Kea we are
privileged to be guests.  Without their kind hospitality, the Keck
observations presented here would not have been possible.  The authors
have made use of the SIMBAD database, the Vienna Atomic Line Database,
and NASA's Astrophysics Data System.

This research is made possible by the generous support of Sun
Microsystems, NASA, and the NSF.  SER was supported by the National
Science Foundation Graduate Research Fellowship.  GL received support
from the NSF Career grant (No. 0449986).  SSV's work was supported by
the NSF grant AST-0307493.  DAF was supported by Research Corporation's
Cottrell Science Scholar program and by NASA grant NNG05G164G.  We thank
the Michelson Science Center for travel support through the KDPA
program.

{\it Facilities:} \facility{Keck I (HIRES)}, \facility{APT}

\clearpage

\begin{deluxetable}{lll}
\tablenum{1}
\tablecaption{Stellar Parameters}
\tablewidth{0pt}
\tablehead{\colhead{Parameter}  & \colhead{HD 5319} & \colhead{HD 75898} \\
}
\startdata
V                      &    8.05             &   8.03               \\
$M_V$                  &    3.05             &   3.49               \\
B-V                    &    0.985            &   0.626              \\
Spectral Type          &    G5 IV            &   G0 V               \\
Distance (pc)          &    100.0            &   80.58              \\
$L_{\star}$ (\lsun)    &    4.6		     &   3.0		    \\
${\rm [Fe/H]}$         &    0.15 (0.05)\tablenotemark{a}      &
                                                 0.27 (0.05)        \\
$T_{\rm eff}$ (K)       &    5052 (50)        &   6021 (50)          \\
\vsini \ks             &    3.31 (0.50)      &   4.54 (0.50)        \\
\logg                  &    3.57 (0.15)      &   4.16 (0.15)        \\
$M_{\star}$ (\msun)     & (1.38) 1.56 (1.74)\tablenotemark{b}  & 
                                                 (1.15) 1.28 (1.41) \\
$R_{\star}$ (\rsun)     & (2.85) 3.26 (3.76)  &   (1.42) 1.6 (1.78)  \\
Age (Gyr)	       & (1.72) 2.40 (3.60)  &   (3.00) 3.80 (5.60) \\
\shk                   &   0.12              &   0.15               \\
\rhk                   &   -5.34             &   -5.02              \\
\prot (d)              &   19.0 d            &   12.6 d             \\
$\sigma_{phot}$ (mag)  & 0.0017              &   \nodata            \\
\enddata
\tablenotetext{a}{Numbers in parentheses give $1\sigma$ uncertainties of
atmospheric parameter measurements.}
\tablenotetext{b}{Double set of parentheses indicates 
95\% confidence intervals for stellar masses and radii.}
\end{deluxetable}
\clearpage

\clearpage
\begin{deluxetable}{rrr}
\tabletypesize{\footnotesize}
\tablecolumns{3}
\tablenum{2}
\tablewidth{0pt}
\tablecaption{Radial Velocities for 5319}
\tablehead{\colhead{JD - 2440000}                            &
           \colhead{RV }                                     &
           \colhead{Uncertainty }                           \\
           \colhead{ }                                       &
           \colhead{\ms}                                     &
           \colhead{\ms}                                    \\
}
\startdata
  13014.75563 &      11.18 &    3.64  \\
  13015.76065 &      20.26 &    3.43  \\
  13016.76506 &      16.41 &    2.94  \\
  13191.11010 &     -46.62 &    2.25  \\
  13207.07543 &     -36.26 &    1.86  \\
  13208.06565 &     -40.49 &    2.10  \\
  13367.70779 &     -19.79 &    1.62  \\
  13368.71602 &     -24.87 &    1.56  \\
  13369.72525 &     -21.66 &    1.49  \\
  13397.71935 &      -1.62 &    1.61  \\
  13694.76591 &      20.56 &    1.60  \\
  13695.77109 &      31.27 &    1.66  \\
  13696.74616 &      31.53 &    1.68  \\
  13724.77909 &      14.52 &    1.56  \\
  13750.73625 &      18.19 &    1.57  \\
  13775.72005 &       7.61 &    1.82  \\
  13776.70605 &       3.29 &    1.81  \\
  13777.72076 &       1.23 &    2.08  \\
  13778.71684 &     -11.90 &    1.83  \\
  13779.74125 &       5.89 &    1.93  \\
  13927.04846 &     -19.25 &    1.70  \\
  13933.04490 &     -23.28 &    1.87  \\
  13959.09166 &     -16.17 &    2.05  \\
  13961.03675 &     -13.30 &    1.77  \\
  13961.04020 &     -13.63 &    1.88  \\
  13981.90601 &     -20.45 &    1.87  \\
  14023.77602 &     -10.05 &    1.92  \\
  14083.83368 &       1.41 &    1.97  \\
  14085.90265 &      10.07 &    1.99  \\
  14129.77461 &      22.85 &    1.89  \\
\enddata 
\end{deluxetable}
\clearpage

\clearpage
\begin{deluxetable}{lll}
\tablenum{3}
\tablecaption{Orbital Parameters}
\tablewidth{0pt}
\tablehead{\colhead{Parameter}  & \colhead{HD 5319} & \colhead{HD 75898} \\
} 
\startdata
P (d)                &   674.6  (17)\tablenotemark{a}    &    418.2 (5.7)        \\
${\rm T}_{\rm p}$ (JD-2440000)   &   13067.7 (77)       &    12907.0 (37)      \\
$\omega$ (deg)           &   76.3 (35)          &    263.7 (30)           \\
ecc                      &   0.12 (0.08)        &    0.10 (0.05)        \\
K$_1$ (\ms)              &   33.6 (4.3)         &    58.2 (3.1)           \\
$dv/dt$ (\ms d$^{-1}$)   &   0.0249 (0.0040)    &    -0.0400 (0.0056)
\\
$a$ (AU)                 &   1.75                &    1.19             \\
$a_1 \sin i$ (AU)        &   0.00207             &    0.00222            \\
f$_1$(m) (M$_\odot$)     &   2.37e-09            &    8.39e-09            \\
$M\sin i$ (M$_{Jup}$)    &   1.94                &    2.51               \\
${\rm Nobs}$             &   30                  &    20                 \\
Assumed jitter (\ms)     &   4.6                 &    2.6                \\
RMS (\ms)                &   6.08                &    5.48               \\
Reduced \chisq           &   1.22                &    1.77               \\
FAP                      &   $0.0013$           &    $< 0.0001$          \\
\enddata                        
\tablenotetext{a}{Uncertainties of Keplerian orbital parameters are
given in parentheses.}
\end{deluxetable}                
\clearpage

\clearpage
\begin{deluxetable}{rrr}
\tabletypesize{\footnotesize}
\tablecolumns{3}
\tablenum{4}
\tablewidth{0pt}
\tablecaption{Radial Velocities for 75898}
\tablehead{\colhead{JD - 2440000}                            &
           \colhead{RV }                                     &
           \colhead{Uncertainty }                           \\
           \colhead{ }                                       &
           \colhead{\ms}                                     &
           \colhead{\ms}                                    \\
}
\startdata
  13014.94581 &      74.84 &    2.11  \\
  13015.94526 &      70.75 &    1.95  \\
  13016.95038 &      75.31 &    2.22  \\
  13071.85048 &      49.86 &    2.47  \\
  13398.05874 &      48.47 &    1.48  \\
  13480.81249 &      33.98 &    1.85  \\
  13724.06020 &     -41.40 &    1.47  \\
  13747.02281 &     -15.98 &    1.70  \\
  13747.95543 &     -12.07 &    1.75  \\
  13748.91982 &      -3.75 &    1.73  \\
  13749.85611 &     -12.29 &    1.31  \\
  13750.86351 &     -22.30 &    1.64  \\
  13752.96966 &     -11.73 &    1.27  \\
  13775.92652 &      15.00 &    1.48  \\
  13776.94321 &       8.26 &    1.88  \\
  13841.81102 &      44.55 &    1.91  \\
  14083.98611 &     -84.77 &    1.39  \\
  14084.91551 &     -80.23 &    1.47  \\
  14086.10668 &     -89.70 &    1.45  \\
  14129.98648 &     -69.14 &    1.52  \\
\enddata 
\end{deluxetable}
\clearpage

\clearpage
\begin{deluxetable}{lrll}
\tablenum{5}
\tablecaption{Constraints on orbits of implied long-period components}
\tablewidth{0pt}
\tablehead{\colhead{Object Type} & \colhead{} & \colhead{HD 5319 c} &
  \colhead{HD 75898 c} \\} 
\startdata
Star & & & \\
& \msini (M$_{\odot}$)     &    0.08--0.65\tablenotemark{a}    &    0.08--0.60       \\
& $a$ (AU)                 &   22--630\tablenotemark{b}  &    17--470          \\
& $P$ (yr)                 &   81--10600           &    58--7400          \\
Brown Dwarf & & & \\
& \msini (M$_{Jup}$)       &   13--83.8           &     13--83.8          \\
& $a$ (AU)                 &   22--630            &    17--470           \\
& $P$ (yr)                 &   24--2045           &    18--1900           \\
Planet & & & \\
& \msini (M$_{Jup}$)       &   1.0--13            &     1.0--13           \\
& $a$ (AU)                 &   4.4--85            &    3.0--65           \\
& $P$ (yr)                 &   7.3--24            &    4.5--18           \\
\enddata                        
\tablenotetext{a}{Each entry in this table gives the minimum and maximum
values a particular parameter can assume, according to the measured
center-of-mass acceleration and stability constraints.}
\tablenotetext{b}{Maximum periods and semimajor axes for each object
class are calculated assuming $e_{\rm max} = 0.8$ (see \S 4.1 for
discussion).}
\end{deluxetable}                
\clearpage

\clearpage

\begin{figure}
\epsscale{0.60}
\plotone{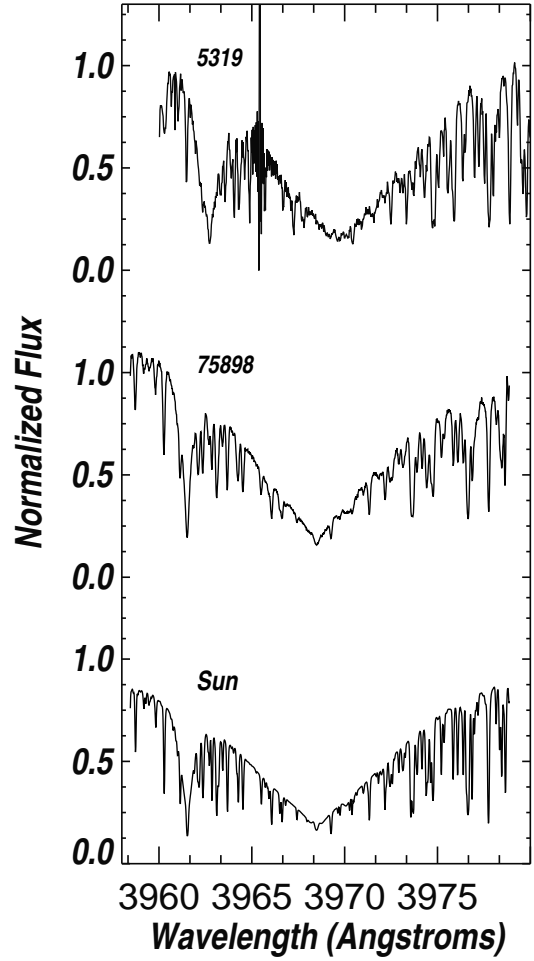}
\caption{Ca II H line for HD 5319 and HD 75898, with the same wavelength
segment of the solar spectrum shown for comparison. Both of these stars
are chromospherically inactive, without emission in the cores of Ca
II H\&K.}
\label{caHK}
\end{figure}
\clearpage    

\begin{figure}
\plotone{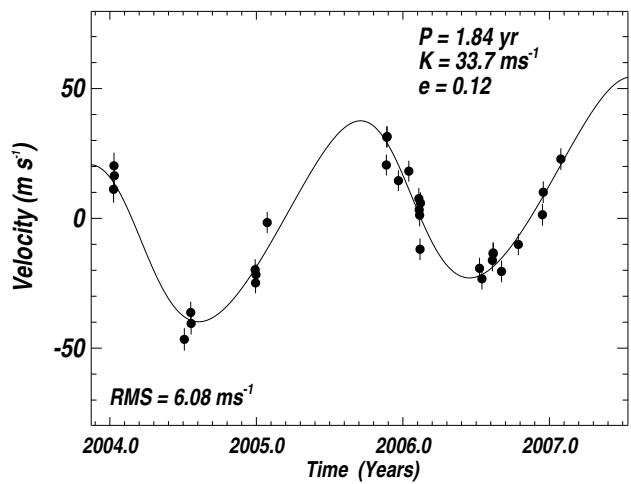}
\figcaption{Radial velocities for HD 5319.  The velocity error bars show
the single measurement precision listed in Table 2, added in quadrature
with 4.6 \ms stellar jitter.  \chisq = 1.22 for a Keplerian fit plus a
constant center-of-mass acceleration of 9.11~\ms~yr$^{-1}$.  Assuming a
stellar mass of 1.56 \msun, we derive a planet mass \msini = 1.94 \mjup,
and a semi-major axis $a = 1.75$~AU.}
\label{rv5319}
\end{figure}
\clearpage
    
\begin{figure}
\plotone{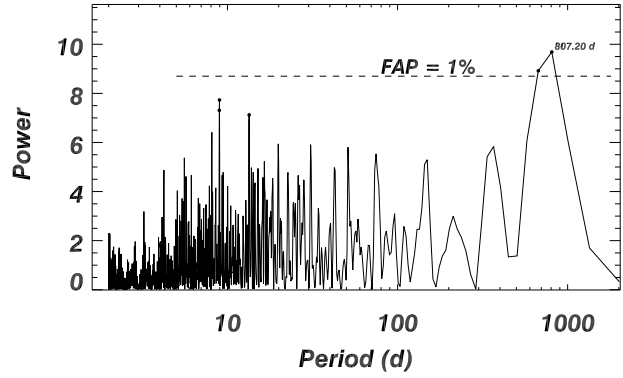}
\figcaption{Periodogram of HD 5319 radial velocities.  The strong peak
spanning 600-900 days has FAP = 0.0013, giving a $99.87\%$ probability
that this peak has a physical, non-noise source.}
\label{period5319}
\end{figure}
\clearpage

\begin{figure}
\plotone{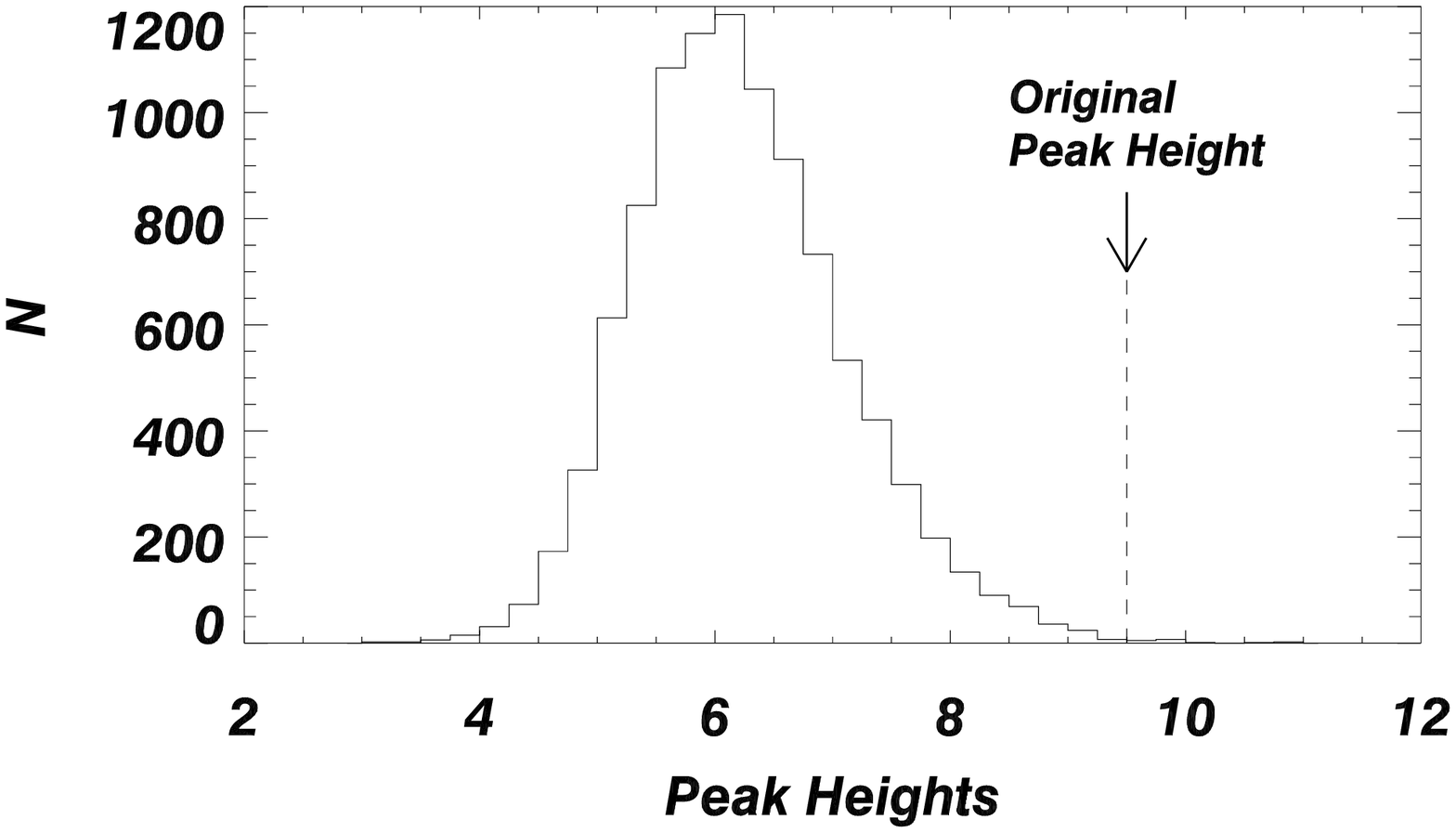}
\figcaption{FAP determination for HD 5319.  Histogram shows maximum
periodogram peak heights in 10,000 synthetic RV data sets, selected with
replacement from the measured radial velocities.  Only 13 trials yielded
maximum power greater than the original periodogram, for ${\rm FAP} =
0.0013$.}
\label{fap5319}
\end{figure}
\clearpage
    
\begin{figure}
\plotone{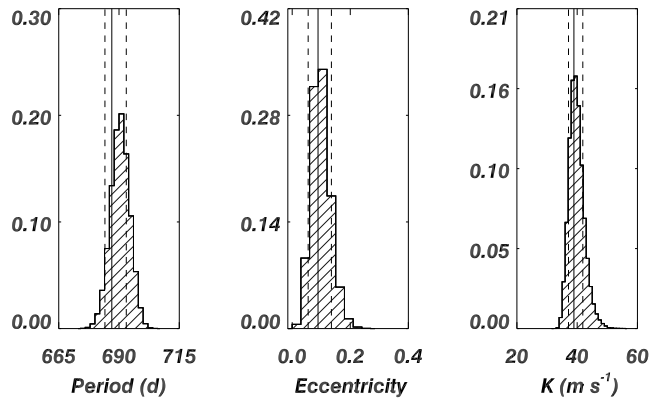}
\figcaption{Markov Chain Monte Carlo models of the radial velocity data
for HD 5319 produce posterior distributions for the orbital period,
eccentricity, and radial velocity semi-amplitude.  The mean values of
each MCMC histogram are broadly consistent with the Keplerian parameters
determined by the Levenberg-Marquardt algorithm, and show convergence to
a single set of orbital parameters.}
\label{hist5319}
\end{figure}

\begin{figure}
\plotone{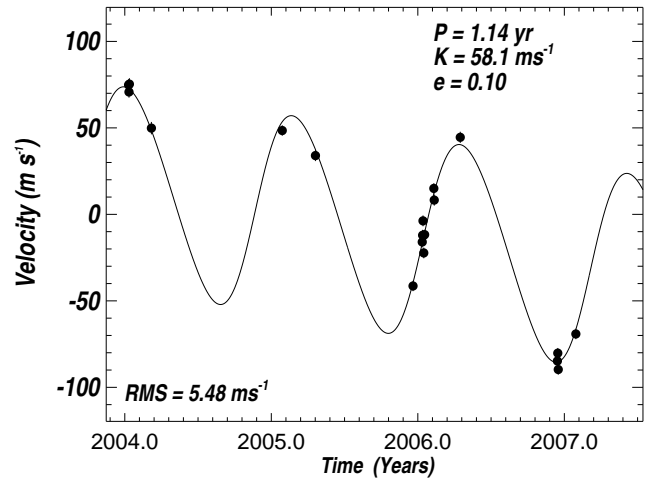}
\figcaption{Radial velocities for HD 75898.  The velocity error bars
show the single measurement precision given in Table 2, added in
quadrature with 2.6 \ms stellar jitter.  This gives \chisq = 1.77 for
the Keplerian fit, with a residual linear trend of -14.6~\ms~yr$^{-1}$.
Assuming a stellar mass of 1.28 \msun, we derive a planet mass \msini =
2.51 \mjup, and semimajor axis $a = 1.19$~AU.}
\label{rv75898}
\end{figure}
\clearpage

\begin{figure}
\plotone{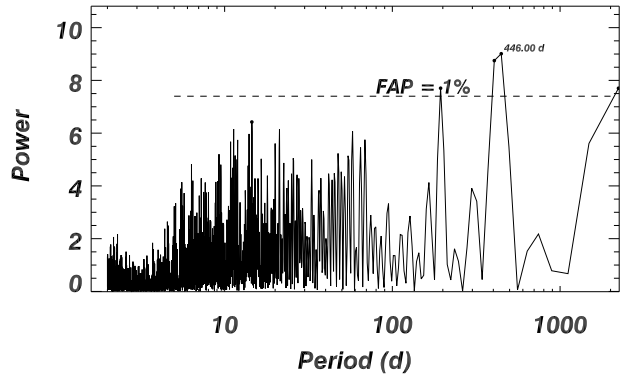}
\figcaption{Periodogram of HD 75898 radial velocities.  The peak at 446
days has ${\rm FAP} < 0.0001$.  The 200-day peak is an alias of the
true, $\sim 400$ day period; the artifact of the seven-month observing
season for this star near the ecliptic.  The observations in late 2006
and early 2007 break this alias and rule out a 200-day period for HD
75898~b.  The rising power toward 2000 days, twice our observational
baseline, is the first hint of a third component of the HD 75898 system
(see \S 4 for details).}
\label{per75898}
\end{figure}
\clearpage

\begin{figure}
\plotone{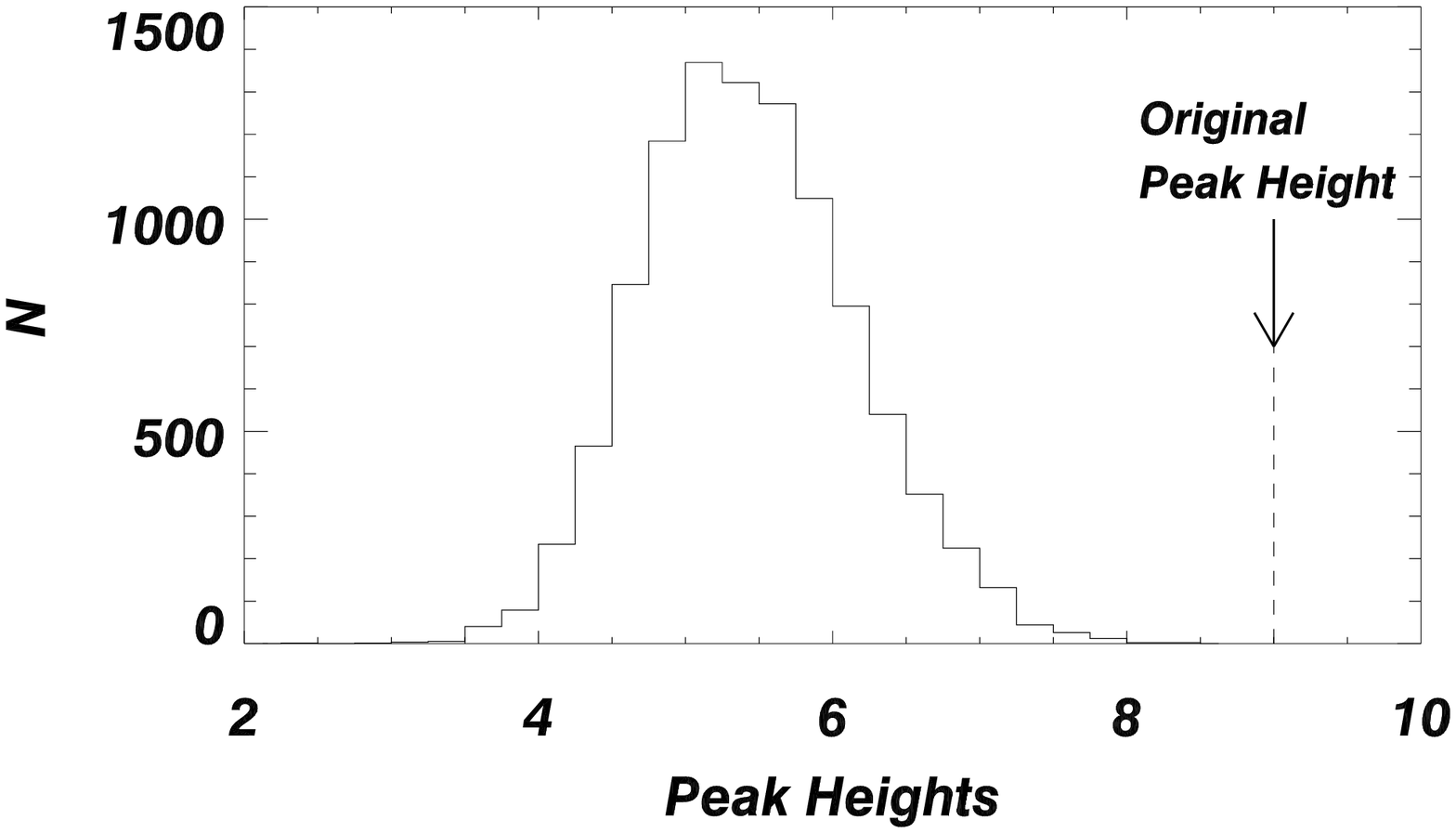}
\figcaption{FAP determination for HD 75898.  Histogram shows maximum
periodogram peak heights in 10,000 synthetic data sets, selected with
replacement from the measured radial velocities.  No trial yielded
maximum power greater than the original periodogram, for ${\rm FAP} <
0.0001$.}
\label{fap75898}
\end{figure}
\clearpage

\begin{figure}
\plotone{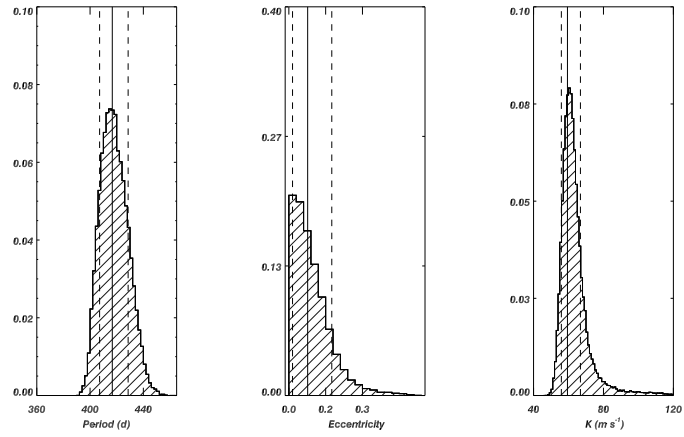}
\figcaption{Markov Chain Monte Carlo posterior distributions for the
period, eccentricity, and radial velocity semiamplitude of HD 75898.
The mean of each distribution is well matched with the corresponding
result from our Levenberg-Marquardt analysis.}
\label{hist75898}
\end{figure}
\clearpage

\begin{figure}
\plotone{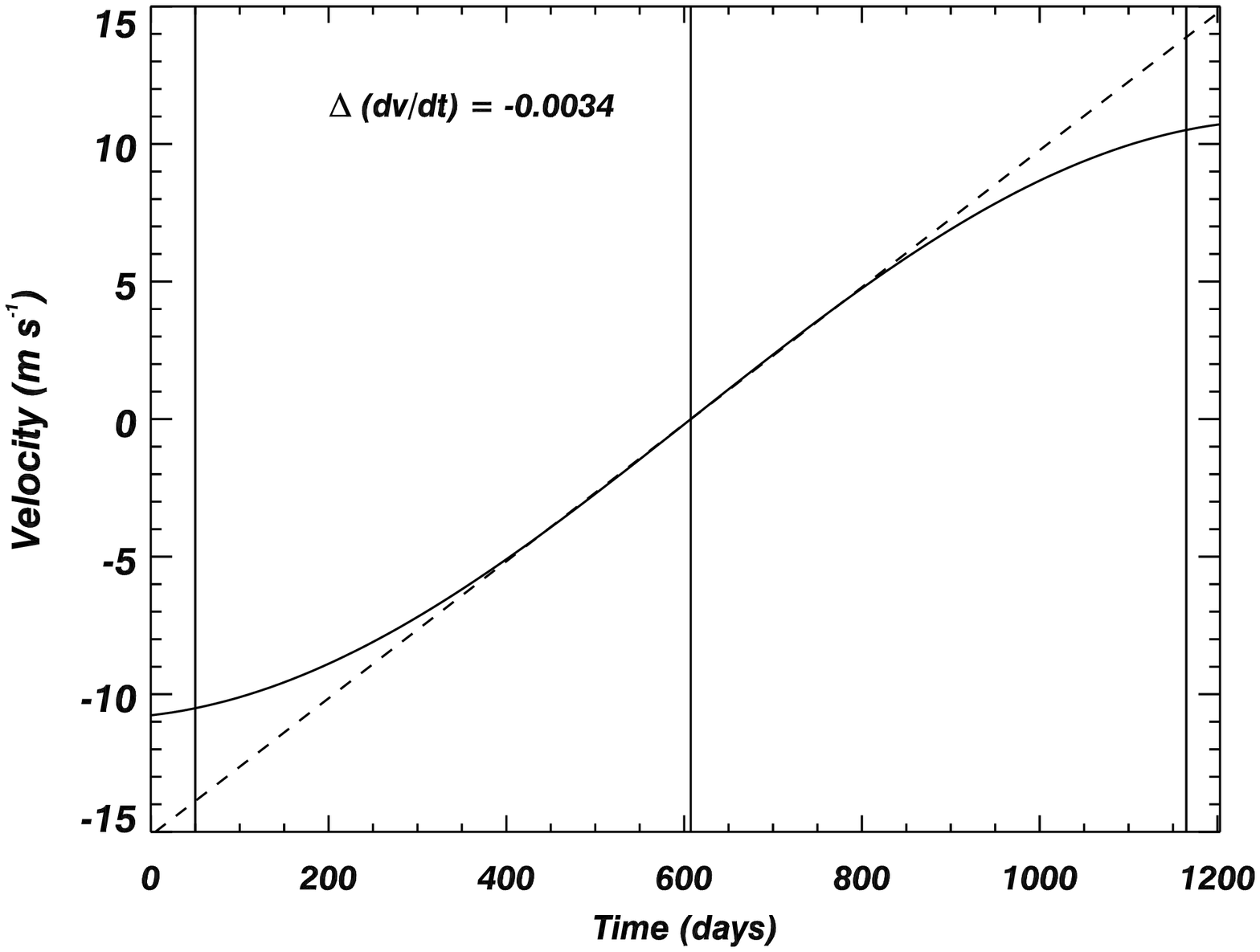}
\figcaption{Theoretical radial velocity curve of HD 5319~c at its
smallest possible mass and period, \msini = 1.0 \mjup and $P = 7.3$ yr.
Vertical lines denote the beginning, midpoint and end of our
observational baseline.  The dashed line shows the center-of-mass
acceleration, 0.0249 \ms~day$^{-1}$.  The difference between the linear
slope of this theoretical RV curve and the measured $dv/dt$,
$|\Delta(dv/dt)| = 0.0034$, is less than the uncertainty in $dv/dt$
(0.0040 \ms~day$^{-1}$.}
\label{hd5319c}
\end{figure}

\begin{figure}
\plotone{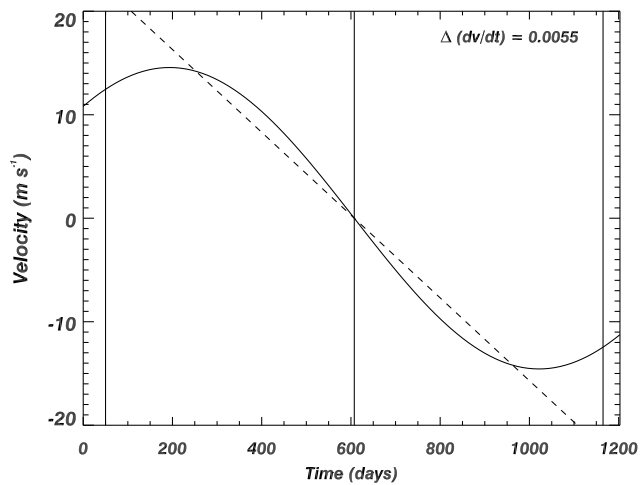}
\figcaption{Theoretical radial velocity curve of HD 75898~c at its
shortest possible period, $4.5$~years, and smallest mass, $1 M_{\rm
Jup}$.  Vertical lines denote the beginning, midpoint and end of our
observational baseline.  The dashed line shows the center-of-mass
acceleration, -0.0400 \ms~day$^{-1}$.  The difference between the linear
slope of this theoretical RV curve and the measured $dv/dt$,
$|\Delta(dv/dt)| = 0.0055$, is consistent with the measured uncertainty
$\sigma(dv/dt) = 0.0056$ \ms~day$^{1}$.}
\label{hd75898c}
\end{figure}


\end{document}